# Simple model for the static structure and the mean coordination of amorphous solids

*Alessio Zaccone*

Institute for Chemical and Bioengineering,

Department of Chemistry and Applied Biosciences,

ETH Zurich,

8093 Zurich, Switzerland.

CORRESPONDING AUTHOR

Alessio Zaccone

Email: alessio.zaccone@chem.ethz.ch.

Fax: 0041-44-6321082.

**ABSTRACT**

We propose a simple route to evaluate the static structure, in terms of average coordination, of completely disordered solids with spherical constituents, from ca. 55% volume fraction up to random close packing, in the absence of structural heterogeneities. Based on the current understanding, according to which the structure-determining interaction in amorphous solids is the hard-core repulsion while weaker, longer-range interactions are mere perturbations, the model yields the average coordination in the solid as a result of a hyperquenching process where the instantaneous structure of the precursor liquid snapshot is distorted to the same degree required to quench the hard-sphere liquid into the isostatic jammed state at $\phi = 0.64$. The characteristic length of distortion turns out to be about 3% of the particle diameter. Extrapolating to lower volume fractions, this is thus the quenching route leading to the most spatially homogeneous states. Thus the model can be usefully employed to quantitatively assess the degree of structural inhomogeneity in amorphous solids. When spatial inhomogeneity is small, as for very dense systems, the model can be used to evaluate coordination-dependent macroscopic properties (e.g. the elastic moduli) as shown in parallel works.

Dynamical arrest in condensed and soft condensed matter represents an open problem where dynamical and structural heterogeneities are thought to play an important if not crucial role. To date, besides experimental and simulation techniques, there are no simple quantitative methods available to analyze and rationalize the origin and extent of structural heterogeneities in amorphous solids. In particular, it lacks a well-defined term of comparison for a quantitative assessment of structural inhomogeneity which makes it impossible to systematically investigate the relation of the latter to microscopic parameters such as e.g. the interaction potential between particles. What is commonly accepted, and observed, is that at high volume fraction (density) the structure is entirely determined by the mutual (hard-sphere) impenetrability between particles which results in a spatially homogeneous distribution of the building blocks. The situation does not change by adding an even strong attractive component of interaction which represents a mere, weak perturbation. Upon lowering the volume fraction $\phi$, below $\phi \approx 0.5$, local crowding of particles is favoured by attraction, which also becomes largely responsible for the mechanical stability, leading to clustering.

Furthermore, there are no means available to a priori estimate the short-range structure of amorphous solids (gels, glasses), which in many cases is affected by structural inhomogeneity, as a function of the volume fraction. This precludes any systematic description of the macroscopic properties of disordered solids.

In an effort to arrive at a more quantitative and directly applicable criterion for structural inhomogeneity in amorphous solids, we propose a method to evaluate the short-range structure in terms of mean coordination corresponding to the highest degree of structural homogeneity achievable upon fast quenching from the liquid state. This provides a baseline for the evaluation of structural inhomogeneity in terms of an observable quantity such as the mean coordination. Further, in systems at high density, where structural inhomogeneity is small, the model provides a way to a priori estimate the mean coordination as a function of volume fraction, which can be used in the evaluation of coordination-dependent macroscopic properties such as elasticity.

It is common believe that a rapid enough quench of a supercooled liquid is able to cause the freezing-in of the liquid constituents (atoms, molecules) almost instantaneously, so that the resulting (solid) glassy state presents a spatial organization which cannot be distinguished, in practice, from that of the liquid snapshot at the quenching time. Following Alexander [1], this corresponds to the assumption that the set of interparticle distances in the *rigid* reference glass state, $\{\mathbf{R}\}$, is the same or approximately the same as the set of interparticle distances in the liquid snapshot, $\{\mathbf{r}\}$, at the quenching time $t_q$ (which is the instant at which the quench is switched on)

$$\{\mathbf{R}\}_{glass} \approx \{\mathbf{r}(t_q)\}_{liquid} \qquad (1)$$

In reality the situation is much more complicated than this simplistic picture because the atomic positions in the liquid snapshot correspond to an unstable structure which is out of mechanical equilibrium.[1] Even for an ideal, infinitely fast quench, where any diffusive motion is suppressed, a solid-like relaxation process accompanies the quench and is responsible for the creation, through a hierarchy of restructuring and buckling phenomena, of the mechanically stable structure which characterizes the final amorphous solid.[1] The simplest case one can think of is that of a hard-sphere liquid where the interaction between particles reduces to mutual impenetrability at contact. Upon quenching (at zero applied pressure) there is a unique final state which is mechanically stable and is given by the (maximally random) jammed state of hard spheres where (marginal) geometric rigidity is ensured by the (Maxwell) isostatic condition $z = 2d = 6$, expressing the equality between geometric constraints and degrees of freedom. This happens at the volume fraction of random close packing $\phi \approx 0.64$, while at lower volume fractions the quenched metastable states are usually very fragile, unless an attractive interaction is present, since the elasticity then is due exclusively to particle caging. If, on the other hand, an attractive interaction is present, mechanically stable states can be obtained even at $\phi \ll 0.64$. In the latter case, however, the extent of attractive interaction also affects the spatial organization so that a strong short-range attraction may favour the insurgence of structural heterogeneity and a higher probability of finding particles near contact (which manifests itself in higher nearest-neighbour peaks in the radial distribution function). Upon lowering the volume fraction below 0.5-0.4 this effect becomes evident giving rise to the typically observed structural heterogeneity in colloidal gels mainly in form of clusters.

Thanks to the unique character of the jamming point of hard-spheres which corresponds to a well-defined (critical) point in the $z - \phi$ plane,[2] it is possible to analytically estimate the extent of rearrangements in the quenching process by which a dense hard-sphere liquid at $\phi = 0.64$ is quenched into a mechanically stable (maximally random) jammed solid with $z = 2d$. Since the jammed state of purely hard-spheres, with no attractive component of interaction, is the most mechanically stable state at zero pressure (of course with soft spheres and $p > 0$ more mechanically stable structures can be created though this case will not be considered here), the degree of restructuring during the quench leading to the jammed state represents a minimum. Extrapolating this result to lower volume fractions gives the coordination of the amorphous solids which at that volume fraction possess the highest degree of spatial homogeneity. The evolution of the coordination with increasing volume fraction, as calculated in this way from the integral equation theory of liquids, purely reflects the geometric random correlations between particles.

The degree of restructuring upon quenching will be calculated in terms of the interparticle distance at which particles, which will form permanent contacts with the given one upon quenching, find themselves in the liquid snapshot.

The structure of a (disordered) fluid of $N$ atoms is entirely described via the $n$-particle distribution function, defined by

$$g_N^{(n)}(\mathbf{r}^n) = \frac{\rho_N^{(n)}(\mathbf{r}_1,...,\mathbf{r}_n)}{\prod_{i=1}^{N}\rho_N^{(1)}(\mathbf{r}_i)} \qquad (2)$$

For a uniform system $\rho_N^{(1)}(\mathbf{r}) = N/V = \rho$, i.e. the number density of the fluid. If the system is homogeneous and isotropic, we have $g_N^{(2)}(\mathbf{r}_1,\mathbf{r}_2) = g(r)$, i.e. the radial distribution function (rdf). The delta function representation of the rdf gives clear evidence of its geometrical meaning in terms of correlation between atoms positioned around a reference one[3]

$$\rho g(r) = \frac{\rho^2}{N}\int g_N^{(2)}(\mathbf{r},\mathbf{r}')d\mathbf{r}' = \left\langle \frac{1}{N}\sum_{i=1}^{N}\sum_{j=1}^{N}\delta(\mathbf{r}-\mathbf{r}_j+\mathbf{r}_i) \right\rangle \qquad (3)$$

From this definition, the concept of coordination shell (in analogy with crystalline solids) is immediately evident, as we shall see also below. The direct correlation function, $c(r)$, is defined by the Ornstein-Zernike convolution relation $h(r) = c(r) + \rho\int h(r')c(|\mathbf{r}-\mathbf{r}'|)d\mathbf{r}'$ which can be solved within the Percus-Yevick closure approximation, $c(r) \approx [1-e^{\beta U(r)}]g(r)$, the solution being piecewise analytic.[3] Since we are interested in the near-contact region, we can approximate the PY solution in the range $1 \leqslant r/\sigma \lesssim 1.1$ by expanding at $r \approx \sigma$

$$g'(r/\sigma';\phi') = (1+\phi'/2)/(1-\phi')^2 - (9/2)\phi'(1+\phi')/(1-\phi')^3 \left(r/\sigma'-1\right) \qquad (4)$$

Correcting for the contact-value at high density and for the phase shift according to the Verlet-Weis semi-empirical prescription, leads to[3]

$$g(r/\sigma;\phi) = g(r/\sigma';\phi') + \delta g_1(r/\sigma) \qquad (5)$$

where $\phi' \approx \phi - \phi^2/16$ and $\sigma' = \sigma(\phi'/\phi)^{1/3}$. The short-range term can be expressed as:

$$\delta g_1(r/\sigma) = \delta g_1(x) = \frac{A}{x}\exp[-\alpha(x-1)]\cos[\alpha(x-1)] \qquad (6)$$

where $x = r/\sigma$, and $A$ and $\alpha$ are coefficients which are only functions of $\phi$ and contain the contact value of the rdf, $g(1;\phi)$. The advantage of this formulation is that the contact behaviour of the PY solution can be modulated by setting $g(1;\phi)$ equal to the value predicted by equations of state valid in the high-density regime, such as the Carnahan-Starling, as well as by numerical simulations. Equations (4)-(6), as it can be easily verified, perfectly agree with the most recent "exact" rdf from computer simulations in the range $1 \leqslant x \lesssim 1.1$.[4] However, it can be noted from the above equations that the rdf is given exclusively as a function of geometrical parameters, independent of temperature (as a consequence of absence of energy parameters in the hard-sphere potential).

The integral of $\rho g(r)$ over a volume element $d^3r$ is just the number of atoms in that volume element and volume integration yields the total number of atoms minus the one at the origin, $\int \rho g(\mathbf{r})d^3r = N-1$. For isotropic systems

$$z = \rho\int_\sigma^r 4\pi r^2 g(r)dr = 24\phi\int_1^x x^2 g(x)dx \qquad (7)$$

which gives the number of particles $z$ in a shell of thickness $r-\sigma$ (or $x-1$) around a given atom. The calculation of the nearest-neighbours number in liquids is regarded as a somewhat ill-posed problem, in the sense that the choice of the upper integration limit might be arbitrary. If one sets the upper boundary equal to the radial coordinate of the first minimum in $g(r)$, the number of atoms in the first coordination shell calculated in this way is always $\approx 11-12$, regardless $\phi$, the location of the minimum being $\phi$-dependent.[3] This happens because the mean nearest-neighbour distance decreases as $\phi$ increases. On the other hand, if one sets the upper integration boundary equal to some fixed arbitrary value (e.g., $r = 1.5\sigma$), the result might describe how the probability of finding neighbours at that distance from the given atom changes with density, but cannot be regarded as an accurate choice. It is well known that the liquid theory predicts a zero number of nearest-neighbours at contact ($r = \sigma$), a feature that well corresponds to the effect of thermal motion of atoms in a liquid. A liquid snapshot is not mechanically stable from any point of view.[1] Submitting the liquid snapshot to a solid-forming quench, even an infinitely fast one, implies a hierarchy of restructuring and buckling phenomena which create internal stresses and permanent contacts until a mechanically stable structure is formed. The following hyperquenching protocol can be used to quench a hard-sphere liquid at $\phi = 0.64$ into a jammed state:

(i) quench the dense hard-sphere liquid down to $T = 0$ instantaneously such that any diffusive motion is suppressed and the atomic configuration remains as in the liquid snapshot;

(ii) instantaneously increase the particle diameter from $\sigma$ to $\sigma + \varepsilon\sigma$.

What one obtains from step (i) is clearly a configuration corresponding to the liquid snapshot. Then, in step (ii), the reference particle comes into *contact* with a number of nearest neighbours which depends upon the value of $\varepsilon$. Note that since collisions among particles may occur in liquid, at step (i) some particles may be already at contact. This requires displacing the contacted particles by a value of $\varepsilon\sigma$ at step (ii). Such an effect is ignored because the $\varepsilon$ value found in the following is very small. Hence, $\varepsilon$ is the parameter which quantifies the degree of restructuring upon quenching and may be seen also as a characteristic length-scale of distortion upon quenching.

Accordingly, the mean coordination in the so obtained quenched (jammed) state can be calculated as

$$z(\phi) = \rho \int_{\sigma}^{\sigma+\varepsilon\sigma} 4\pi r^2 g(r) dr = 24\phi \int_{1}^{1+\varepsilon} x^2 g(x) dx \qquad (8)$$

A somewhat similar procedure has been used to derive the $\phi$-scaling of the incremental coordination number in the compressive regime of soft repulsive particles (see Eq. 1.4 in Ref. [4]). Note that, in order to preserve $\phi$ throughout the process, the volume of the quenched state must transform as $V(\hat{\sigma}/\sigma)^3 = V(1+\varepsilon)^3$.

The integral in Eq. (8) can be rewritten, in terms of the interparticle gap $l = (r-\sigma)/\sigma = x-1$, as $z = 24\phi \int_0^l (1+l)^2 g(l)dl$. Then, applying Eqs. (4)-(6), one can integrate analytically, to find

$$z(l;\phi) \equiv 24\phi \int_0^l (1+l)^2 g(l)dl = 2\phi \left\{ l \left[ 4\frac{1+\phi'/2}{(1-\phi')^2}(3+3l+l^2) - \frac{9}{2}\phi'\frac{1+\phi'}{(1-\phi')^3}l(6+8l+3l^2) \right] \right.$$
$$\left. +6\frac{A}{\alpha}\exp(-\alpha l)\left[\alpha\exp(-\alpha l) + \sin\alpha l + \alpha(1+l)(\sin\alpha l - \cos\alpha l)\right] \right\} \quad (9)$$

which is accurate results for small gaps ($l < 0.1$). The Verlet-Weis parameters evaluated based on the Carnahan-Starling contact value of the rdf, $g_{CS}(1;\phi')$, are given by

$$A = \frac{3}{4}\frac{\phi'^2(1-0.7117\phi'-0.114\phi'^2)}{(1-\phi')^4}, \qquad \alpha = \frac{18(1-0.7117\phi'-0.114\phi'^2)}{(1-\phi')(1-\phi'/2)} \quad (10)$$

More accurate though longer expressions in the high density fluid branch can be obtained using the Hall equation of state which makes use of a larger number of virial coefficients.[5] At this point, we can determine the value of $\varepsilon$ corresponding to an ideal, infinitely fast quench of the hard-sphere liquid at $\phi = 0.64$ into a mechanically stable jammed solid with $z = 2d = 6$. Using the Hall equation of state for hard-spheres at high-density in the fluid branch, from the latter condition we find

$$\varepsilon \approx 0.032 \quad (11)$$

which is indeed a small number. Hence, the degree of restructuring during the quench into an isostatic mechanically stable jammed state, with no interaction other than the hard-sphere one, amounts to 3% ca. of the particle diameter. This value represents a lower bound. In fact, at lower $\phi$, if no attractive interaction is present which can enforce mechanical stability, the state resulting from hyperquenching will not be rigid because $z<6$ and further rearrangements will be required until a denser configuration with $z=2d=6$ is formed, necessarily implying $\varepsilon > 0.032$. On the other hand, if attractive interactions are present they can ensure mechanical stability due to the formation of initial stresses according to a mechanism that has been proposed and discussed in depth by Alexander.[1] Since, however, attraction leads to an increased spatial heterogeneity, we can regard quenched states with $z$ calculated using Eq. (11) as those having the highest degree of spatial homogeneity. Since spatial heterogeneity is small, because of geometric constraints and hard-core repulsion, at least for $\phi \gtrsim 0.5$, the predictions using Eq. (11) may still be directly applied to a priori calculate the mean coordination in a wide range of practical situations. Indeed, no significant difference in the short-range structure has been detected between attractive and purely hard-sphere colloidal glasses in recent experimental studies at $\phi \approx 0.60$.[6] It is however possible that the approximation may start to lack accuracy upon getting closer to $\phi \approx 0.5$.

We note that Eq. (9) together with (11) provides a relation between $z$ and $\phi$ which can be rather accurately interpolated by a power-law with an exponent of order 4 slightly dependent upon the range of $\phi$. In particular, as one can easily verify, $z \propto \phi^{3.8}$ in the range $0.45 < \phi < 0.64$ with an accuracy of the power-law fitting given by $R^2 = 0.993$, while for $0.55 < \phi < 0.64$ it becomes $z \propto \phi^{4.15}$ with $R^2 = 0.995$.

The method described here can be used to calculate coordination-dependent properties of glassy materials such as the shear modulus, as a function of the volume fraction. In parallel work, in fact, it has been shown that these predictions lead to reasonable result in the case of dense aggregated colloidal systems and attractive colloidal glasses.[7,8] Further, since it provides a simple and systematic way to quantify the mean coordination (thus the short-range structure) of amorphous solids in the absence of structural inhomogeneity, it may also be usefully employed to provide a quantitative criterion in estimating the extent of structural inhomogeneity under different forms of microscopic (attractive) interactions. This may give valuable quantitative information in order to investigate the effect of the nature and strength of the interaction potential on the static structure of glasses, simply by comparison with measured values of the mean coordination in the actual system under study. This will hopefully help to improve our understanding of structural and dynamical heterogeneities and their interplay with dynamical arrest in condensed and soft condensed matter systems.[9]